\documentclass[aps]{revtex4}
\usepackage{times}
\usepackage{epsfig}
\usepackage{amssymb}
\usepackage{bm}
\usepackage{amsmath}
\usepackage{amsfonts}
\usepackage{graphicx}

\begin{document}

\title{On the possibility of observable signatures of $ \mu p $ and $ (\mu ^4\mathrm{He})^{+} $  lines on the spectra of astrophysical sources}

\author{V. Dubrovich$^{1}$, T. Zalialiutdinov$^{2}$}
\affiliation{ 
$^{1}$ Special Astrophysical Observatory, St. Petersburg Branch, Russian Academy of Sciences, 196140, St. Petersburg, Russia \\
$^{2}$ Department of Physics, St. Petersburg State University, Petrodvorets, Ulianovskaya 1, 198504, St. Petersburg, Russia}

\begin{abstract}
We examine the processes of the luminescence in subordinate lines of muonic hydrogen $ \mu p $  and muonic helium ion $ (\mu ^4\mathrm{He})^{+} $ in the presence of background source of X-ray emission. It is supposed that a certain amount of muonic atoms existing in the vicinity of astrophysical source reemits absorbed radiation in the subordinate lines.  The intensity of luminescence of such a process is proportional to the quantum yield which was calculated for different pumping channels and different models of spectra. It is shown that the luminescent lines of muonic hydrogen and muonic helium ion can be very noticeable in the spectrum of background source. 
\end{abstract}

\maketitle

\section{Introduction}
\label{intro}

During the last decade the essential progress in X-ray astronomy has been brought about by the advent of satellite observatories and by the great number of new radio and optical identifications of cosmic X-ray sources \cite{lit1, lit2}. Since the discovery, with rocket-borne instruments, of extrasolar sources of x radiation, it has been clear to most experimenters that a very considerable advance in our knowledge could be obtained with satellite instrumentation. Recent launch of space observatory Spektr-RG reveals new opportunities for observing of cosmic radiation in the energy range of 0.3-30 keV. As a consequence, the calculation of line intensities for different atomic systems and their identification is one of the most important tasks of modern X-ray astronomy \cite{kev1,kev2,dubrPI}. 
 
In the present paper we consider the possibility to observe signatures of muonic hydrogen $ \mu p $ and muonic helium ion $ (\mu ^4\mathrm{He})^{+} $ on the spectrum of cosmic X-ray sources. By cosmic sources of high-energy photons we mean stars, quasars, active galactic nuclei, and etc., in the neighbourhood of which a certain number of muons, protons and alpha particles can be produced.
The intensity of atomic lines resulting from the reemission of high-energy photons absorbed from the source is proportional to the quantum yield. The method used in the present work for the estimations of quantum yield $ q_{ij} $ was first proposed in \cite{bernberndubr} in the context of Cosmic Microwave Background (CMB) distortions and applied later in \cite{dubrlipka, dubr1997} for the calculations of CMB distortions from primary molecules. By definition, the quantum yield $q_{ij}$ is the ratio of the mean number of photons of specified frequency emitted in transition from the upper level $i$ to the lower level $j$ to the number of resonant photons in the pumping channel. Then the luminescence intensity is proportional to the pumping intensity and to the quantum yield $q_{ij}$. Recently the luminescence in primordial helium lines at the prerecombination epoch was considered in \cite{Dubrovich2018,Dubrovich2018-a}. It was shown that luminescent lines can be quite noticeable in the spectrum of blackbody background radiation. In the present paper we extend approach proposed in \cite{Dubrovich2018,Dubrovich2018-a} for the calculations of luminescence intensity in subordinate lines of muonic hydrogen atom $ \mu p $ and muonic helium ion $ (\mu ^4\mathrm{He})^{+} $ in the presence of background source of X-ray emission. 

\section{Formation of luminescent lines of muonic atoms}
\label{method}
The intensity of the luminescent lines is proportional to the quantum yield. The quantum yield can be found from the solution of the system of kinetic balance rate equations \cite{Dubrovich2018}
\begin{eqnarray}
\label{rate1}
\frac{dN_{i}}{dt}=-N_{i}\sum_{j=1}^{K}R_{ij}+\sum_{j=1}^{K}N_{j}R_{ji}
\end{eqnarray}
where $ N_{i} $ is the occupation number of the level $ i $ given by a set of quantum numbers $ n_{i}l_{i}$ ($ n_{i} $ is the principal quantum number, $ l_{i} $ is the orbital quantum number), $ R_{ij} $ is the probability coefficient (transition rate in s$^{-1}$) for the transition $ n_{i}l_{i}\rightarrow n_{j}l_{j}  $, $ t $ is the time, $ K $ is the number of considered bound states. Sum over $ j $ in Eq. (\ref{rate1}) implies summation over the set of quantum numbers $ n_{j}l_{j}$. In the model under consideration, we neglect the angular and spatial distribution of the incident radiation. This is justified by the fact that qualitatively the quantum yield should not change much as the main contribution arises due to spontaneous transitions. Then neglecting the induced emission transition rates in the presence of background radiation are given by the following relations
\begin{eqnarray}
\label{2}
R_{ij}=A_{ij}
,\end{eqnarray}
for the transitions from the upper level to the lower one ($ E_{i}>E_{j} $) and 
\begin{eqnarray}
\label{3}
R_{ij}=\frac{g_{j}}{g_{i}}\frac{c^2}{2h\nu_{ij}^3}I_{\nu}A_{ji}
\end{eqnarray}
for transitions from the lower level to the upper one ($ E_{i}<E_{j} $). Here $ \nu_{ij}=\left|E_{i}-E_{j} \right|/h $ is transition frequency in the laboratory frame of reference, $ A_{ij} $ is the Einstein coefficient for spontaneous transition $ n_{i}l_{i}\rightarrow n_{j}l_{j} $, $ g_{i} $ is the statistical weight of the state $ i $. It is convenient to rewrite Eq. (\ref{rate1}) in terms of Menzel factors \cite{kaplan}
\begin{eqnarray}
\label{menzel}
b_{i}=N_{i}/N^{\rm LTE}_{i},
\end{eqnarray}
where $ N^{\rm LTE}_{i} $ is the equilibrium population of level $ i $ given by Saha equation.
\begin{eqnarray}
\label{saha}
N^{\rm LTE}_{i}=N_{\mu}N_{n}\frac{g_{i}}{4}
\left(\frac{2\pi m_{\mu}k_{B}T_{\rm M}}{h^2}\right)^{-3/2}e^{E_{i}^{\rm ion}/k_{B}T_{\rm M}}.
\end{eqnarray}
Here $ N_{\mu} $ is the muon number density, $ N_{n} $ is number density of atomic nuclei (protons or alpha particles), $ h $ is the Planck constant, $ m_{\mu} $ is the muon mass and $ E_{i}^{\rm ion} $ is the ionization energy for the muonic atom in the state $ i $ ($ E_{1s}^{\rm ion}(\mu p)=2528 $ eV and $ E_{1s}^{\rm ion}(\mu ^4\mathrm{He})^{+}=10942 $ eV) and $ T_{\rm M} $ is the temperture of particles. Then using Eq. (\ref{menzel}) and taking into account that $ N_{i}^{\rm LTE}R_{ij}=N_{j}^{\rm LTE}R_{ji} $ system of rate equations (\ref{rate1}) takes the form \cite{Dubrovich2018,Dubrovich2018-a}
\begin{eqnarray}
\label{rate2}
\frac{db_{i}}{dt}=-\sum_{j=1}^{K}R_{ij}(b_{i}-b_{j}).
\end{eqnarray}
Equation (\ref{rate2}) can be solved analytically \cite{Dubrovich2018}, however in the present work we use numerical solutions. Both methods give the same results. Since we are interested in corrections to equilibrium populations, it is natural to represent  solution of equations (\ref{rate2}) in the form $1 + \Delta b_{i}$. Obviously, the system of equations for corrections $ \Delta b_{i} $ has the same form as for the populations themselves. Therefore below we will understand $ b_{i} $ as corrections to populations. Moreover we need to take into account the probability of muon transition from the state $ i $ to the state $ j $ under the condition that the muon does not decay during its lifetime $ \tau_{\mu}=2.197\times 10^{-6} $ s. This can be done by multiplying each equation of system (\ref{rate2}) by 
branching ratio 
\begin{eqnarray}
\label{p}
p_{ij}=\frac{R_{ij}}{R_{ij}+\tau_{\mu}^{-1}}.
\end{eqnarray}
According to Eqs. (\ref{p}) the absolute probability to emit a photon in transition $ 2p\rightarrow 1s+\gamma(\mathrm{E1}) $  before the natural decay of $ \mu $ is $ p_{2p1s}=0.99 $ both for $ \mu p  $ and $ (\mu ^4\mathrm{He})^{+} $ atoms. The absolute probability to emit two  photons in transition $ 2s\rightarrow 1s+\gamma(2\mathrm{E1}) $ is $ p_{2s1s}=0.003 $ for $ \mu p  $ and $ p_{2s1s}=0.23 $ for $ (\mu ^4\mathrm{He})^{+} $ \cite{nuov,nuov2}. 

By the definition, the quantum yield in the transition between the upper level $ i $ and the lower level $ j $ is the number of uncompensated transitions in this line per one initial excited atom in the pumping line. Since we assumed that the Menzel factor for the upper level of pumping line is $ b_{k}=1 $ at $ t=0 $, then the population of this level is $N_{k}^{\rm LTE}$, i.e., given by Saha equation (\ref{saha}). Finally the number of uncompensated transitions (quantum yield) in line $ i\rightarrow j $ is obtained by multiplying corresponding term in system (\ref{rate2}) by $ N_{i}^{\rm LTE} $
\begin{eqnarray}
\label{yield}
q_{ij}=\frac{N_{i}^{\rm LTE}}{N_{k}^{\rm LTE}}p_{ij}R_{ij}\int\limits_{0}^{\tau_{\mu}}\left(b_{i}(t)-b_{j}(t) \right)dt
.
\end{eqnarray}
Interval of integration over the time $ t $ in Eq. (\ref{yield}) is limited by the muon lifetime $ \tau_{\mu} $, i.e. only transitions that occur in a time shorter than the muon lifetime contribute to the quantum yield. Since the medium is not supposed to be optically thick the Sobolev escape probability is not taken into account in Eq. (\ref{rate2}). 

\begin{figure}
\caption{Quantum yield $ q_{ij} $ in the line $ 3p\rightarrow 1s$ of $ \mu p $ atom for the pumping channel $ 1s\rightarrow 3p $. The bold line corresponds to the blackbody spectrum of the source, see Eq. (\ref{inu1}), dashed and dotted lines correspond to spectrum Eq. (\ref{inu2}) with $ \alpha = 2.5 $ and $ \alpha = 3 $ respectively. Horizontal axis is the photon energy.}
\includegraphics[scale=0.75]{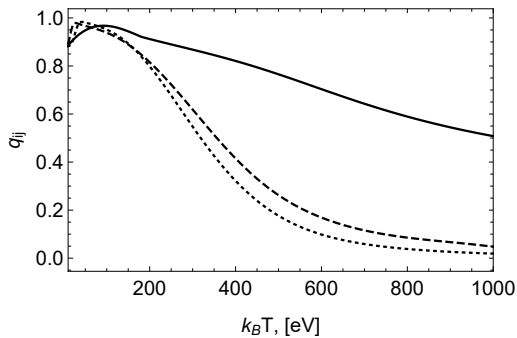}
\label{fig1}
\end{figure}

\begin{figure}[hbtp]
\caption{Quantum yield $ q_{ij} $ in the line $ 3p\rightarrow 2s$ of $ \mu p $ atom for the pumping channel $ 1s\rightarrow 3p $. All notations are the same as for Fig. \ref{fig1}.}
\includegraphics[scale=0.75]{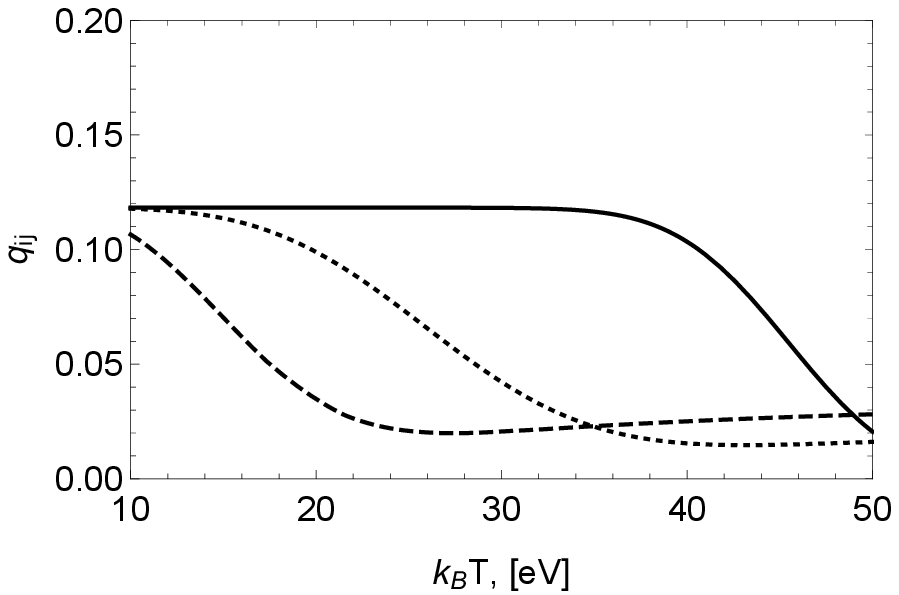}
\label{fig2}
\end{figure}

\begin{figure}[hbtp]
\caption{Quantum yield $ q_{ij} $ in the line $ 4p\rightarrow 1s$ of $ \mu p $ atom for the pumping channel $ 1s\rightarrow 4p $. All notations are the same as for Fig. \ref{fig1}.}
\includegraphics[scale=0.75]{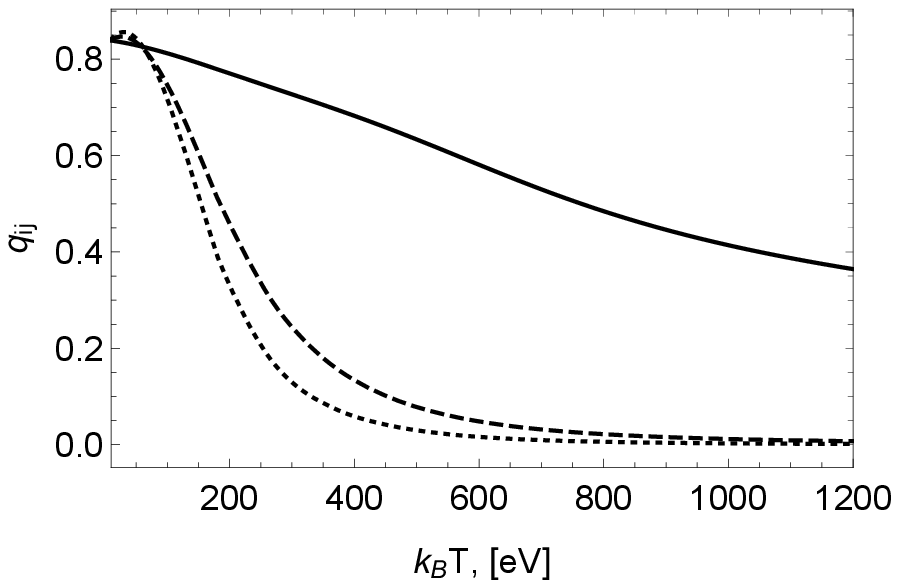}
\label{fig3}
\end{figure}

\begin{figure}[hbtp]
\caption{Quantum yield $ q_{ij} $ in the line $ 4p\rightarrow 2s$ of $ \mu p $ atom for the pumping channel $ 1s\rightarrow 4p $. All notations are the same as for Fig. \ref{fig1}.}
\includegraphics[scale=0.75]{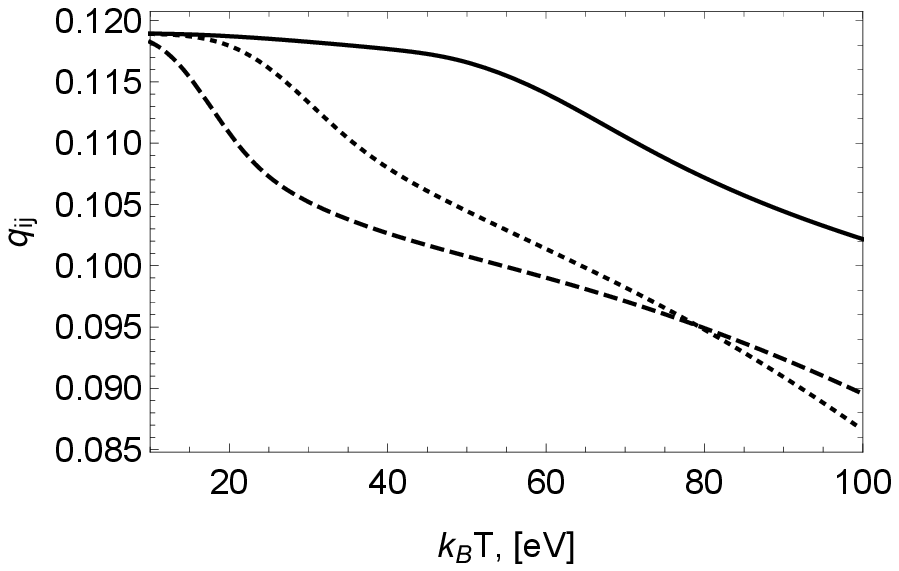}
\label{fig4}
\end{figure}

The absorption rates in Eq. (\ref{rate2}) depend on the radiation intensity $ I_{\nu} $ of source. As an example for our estimations we consider different model of source spectrum. As a first model of spectrum we consider the Planck distribution (in units $\rm Erg \cdot cm^{-2} \cdot s^{-1} \cdot Hz^{-1} \cdot sr^{-1}$)
\begin{eqnarray}
\label{inu1}
I_{\nu}=\frac{2h\nu^3}{c^2}\frac{1}{e^{\frac{hv}{k_BT}}-1}.
\end{eqnarray}
In the second case we consider the power law dependence of the spectrum
\begin{eqnarray}
\label{inu2}
I_{\nu}=\frac{2h\nu^3}{c^2}\left(\frac{k_BT}{h\nu}\right)^{\alpha}
\end{eqnarray}
with $ \alpha=2.5 $ and $ \alpha = 3 $. The choice of the degree of $ \alpha $  is due only to the need to demonstrate how different spectral models work. In our example, we chose two values giving the highest values of quantum yield $ q_{ij} $. Under the real conditions, knowledge of the exact spectrum is necessary \cite{shakura}.
Both spectral models (Eq. (\ref{inu1}) and (\ref{inu2})) depend on effective temperature $ T $ of a source. Within the framework of considered model we will set the radiation temperature of sources $ T= T_{\rm M} $ in Eqs. (\ref{inu1}), (\ref{inu2}). The results of evaluations of Eq. (\ref{yield}) for two different models of spectra are presented in Figs. \ref{fig1}-\ref{fig4}. The pumping channels and considered luminosity lines are chosen so as to get rid of the dependence of the quantum yield on the number densities of states, i.e. we set $ N^{\rm LTE}_{k}=N^{\rm LTE}_{i} $ in Eq. (\ref{yield}).

The calculations are based on the model of a muonic hydrogen and muonic helium ion with $ 27 $ states ($ n\leqslant 10 $ and $ l\leqslant 2 $). Einstein coefficients for the electric dipole transitions were calculated by rescaling of transition rates for ordinary $ \mathrm{H} $ and $ ^4\mathrm{He}^+ $ atoms with the reduced mass $ M_{\rm red}=m_{n}m_{\mu}/(m_{n}+m_{\mu}) $ ($ m_{n} $ and $ m_{\mu} $ are the masses of nuclei and muon respectively) (see also \cite{milotti}). The rates of spontaneous two-photon transition $ 2s\rightarrow 1s+2\gamma(\mathrm{E1}) $ for muonic hydrogen atom and muonic hellium ion are $ A^{2\mathrm{E1}}_{2s1s}(\mu p)=1.53\times 10^{3} $ s$^{-1}$ and $ A^{2\mathrm{E1}}_{2s1s}((\mu ^4\mathrm{He})^{+} )=1.06\times 10^{5} $ s$^{-1}$ respectively \cite{new1,new2,nuov,nuov2}. The probabilities of spontaneous magnetic dipole transition $ 2s\rightarrow 1s+\gamma(\mathrm{M1}) $ are strongly suppressed and not taken into account for both atoms \cite{nuov2}. Numerical calculation of Eq. (\ref{yield}) was carried out both with and without the account for two-photon decay of $2s$ state. It was found that two-photon transition $ 2s\rightarrow 1s+\gamma(2\mathrm{E1}) $ plays negligible role in formation of quantum yield $ q_{ij} $ for both atomic systems. 

The transition energies and Einstein coefficients for $ \mu p $  and $  (\mu ^4\mathrm{He})^{+} $ atoms averaged over orbital momenta are presented in Tables \ref{enandein} and \ref{enandein2} respectively. Averaging over orbital momenta in Tables \ref{enandein} and \ref{enandein2} is performed with the use of equation \cite{weise}
\begin{eqnarray}
\label{avg}
A^{\rm avg}_{n_in_j}=\sum\limits_{l_il_j}\frac{2l_i+1}{n_i^2}A_{n_il_in_jl_j}.
\end{eqnarray}
Solutions of the system Eq. (\ref{rate2}) were checked for convergence with different number $ K $ of considered states. 

\begin{figure}[hbtp]
\caption{Quantum yield $ q_{ij} $ in the line $ 3p\rightarrow 1s$ of $ (\mu \mathrm{He})^{+} $ atom for the pumping channel $ 1s\rightarrow 3p $. All notations are the same as for Fig. \ref{fig1}.}
\includegraphics[scale=0.75]{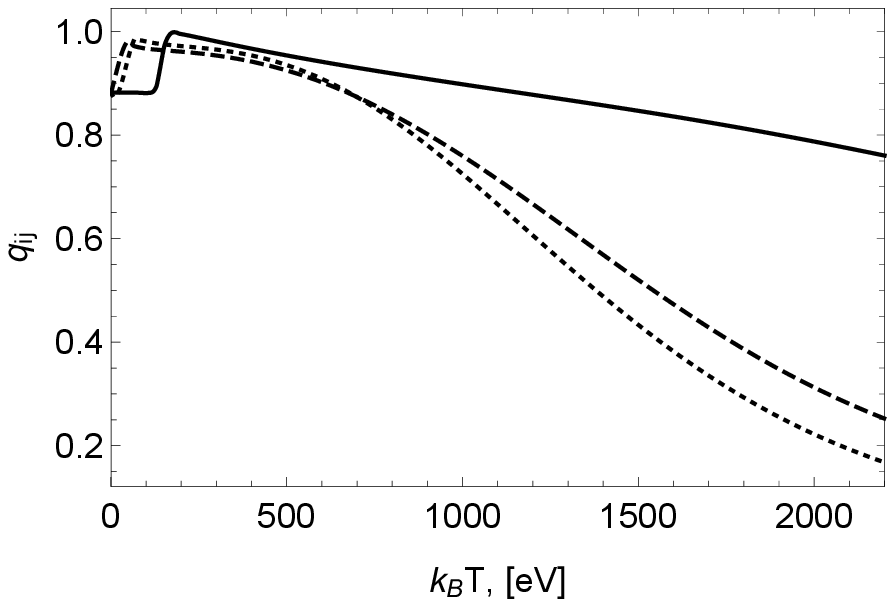}
\label{fig5}
\end{figure}

\begin{figure}[hbtp]
\caption{Quantum yield $ q_{ij} $ in the line $ 3p\rightarrow 2s$ of $ (\mu^4 \mathrm{He})^{+} $ atom for the pumping channel $ 1s\rightarrow 3p $. All notations are the same as for Fig. \ref{fig1}.}
\includegraphics[scale=0.75]{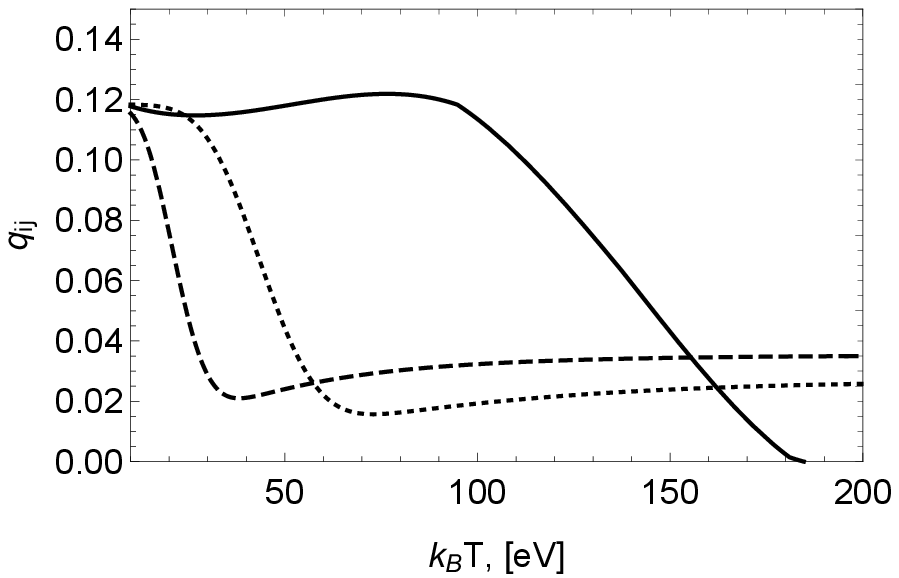}
\label{fig6}
\end{figure}

\begin{figure}[hbtp]
\caption{Quantum yield $ q_{ij} $ in the line $ 4p\rightarrow 1s$ of $ (\mu^4 \mathrm{He})^{+} $ atom for the pumping channel $ 1s\rightarrow 4p $. All notations are the same as for Fig. \ref{fig1}. }
\includegraphics[scale=0.75]{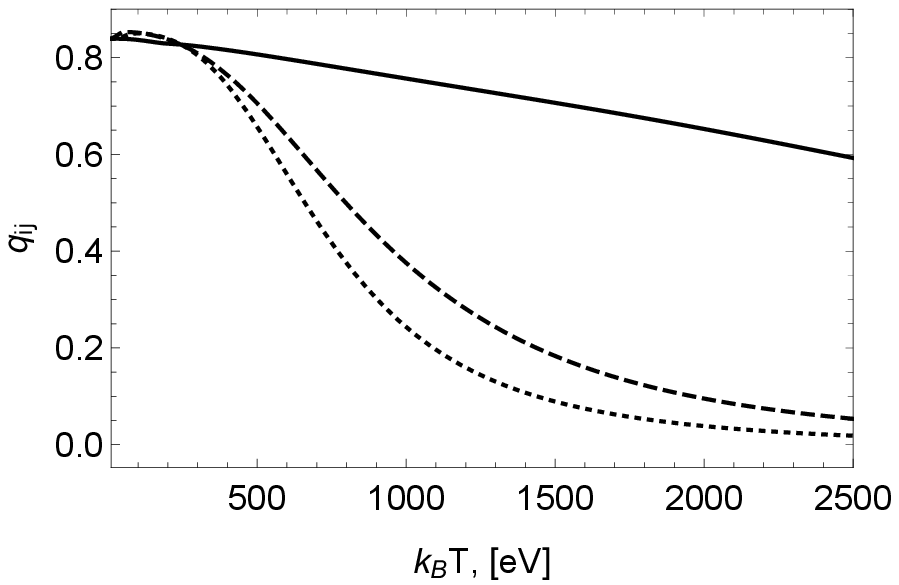}
\label{fig7}
\end{figure}

\begin{figure}[hbtp]
\caption{Quantum yield $ q_{ij} $ in the line $ 4p\rightarrow 2s$ of $ (\mu^4 \mathrm{He})^{+} $ atom for the pumping channel $ 1s\rightarrow 4p $. All notations are the same as for Fig. \ref{fig1}.}
\includegraphics[scale=0.75]{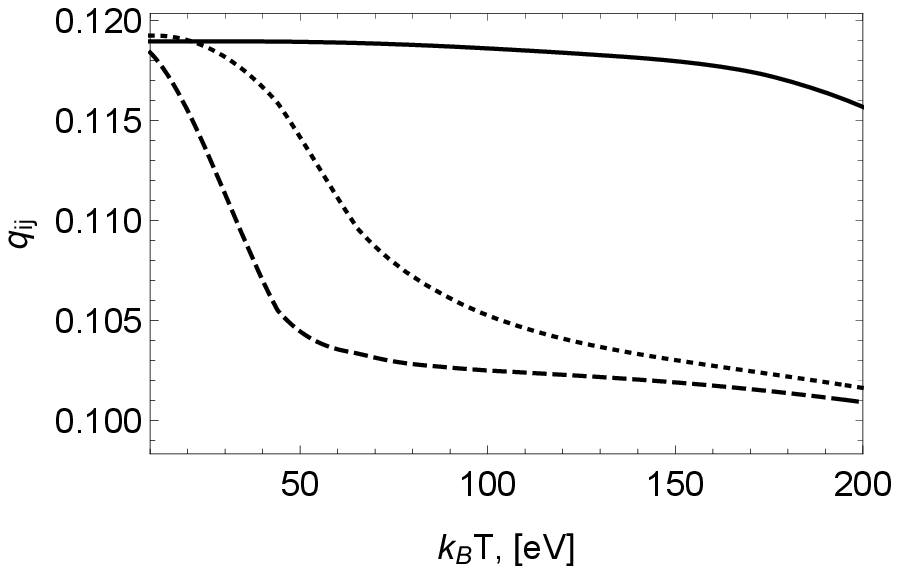}
\label{fig8}
\end{figure}

\begin{table}
\caption{Einstein coefficients for spontaneous emission averaged over angular momenta (see Eq. (\ref{avg})) and transition energies for the first five levels of muonic hydrogen atom $ \mu p $. The first line in each cell indicates Einstein  coefficients $ A^{\rm avg}_{n_in_j} $ (in s$ ^{-1} $) transitions $ n_i\rightarrow n_j $ \cite{milotti}. The second line in each cell indicates the transition energy in eV. }
\begin{tabular}{c p{15mm} p{15mm} p{15mm} p{15mm}}
\hline
$n_i$ $\setminus$ $n_j$   & 1 & 2 & 3 & 4 \\
\hline
2  & 8.737$ \times 10^{10} $ \newline 1896.37 &    &  &        \\
3  & 1.037$ \times 10^{10} $ \newline 2247.55 & 8.200$ \times 10^{9} $ \newline 351.18 & & \\
4  &  2.377$ \times 10^{9} $\newline  2370.46 &  1.565$ \times 10^{9} $ \newline 474.09 & 1.671$ \times 10^{9} $\newline 122.91 & \\
5  &  7.670$ \times 10^{8} $ \newline  2427.35&  4.705$ \times 10^{8} $ \newline 530.98 & 4.092$ \times 10^{8} $\newline 179.80 & 5.019$ \times 10^{8} $ \newline 56.89\\
\hline
\end{tabular}
\label{enandein}
\end{table}

\begin{table}
\caption{Einstein coefficients for spontaneous emission averaged over angular momenta (see Eq. (\ref{avg})) and transition energies for the first five levels of muonic helium ion $ (\mu ^4\mathrm{He})^{+} $. The first line in each cell indicates Einstein coefficients $ A^{\rm avg}_{n_in_j} $ (in s$ ^{-1} $) for transitions $ n_i\rightarrow n_j $. The second line in each cell indicates the transition energy in eV. }
\begin{tabular}{c p{15mm} p{15mm} p{15mm} p{15mm}}
\hline
$n_i$ $\setminus$ $n_j$   & 1 & 2 & 3 & 4 \\
\hline
2  & 1.512$\times 10^{12}$  \newline  8207.04 &   &  & \\
3  & 1.795$\times 10^{11}$  \newline  9726.86 &  1.420$\times 10^{11}$ \newline 1519.82 &  & \\
4  & 4.115$\times 10^{10}$  \newline  10258.8 & 2.710$\times 10^{10}$  \newline   2051.76 & 2.89246$\times 10^{10}$ \newline   531.94 & \\
5  & 1.328$\times 10^{10}$ \newline   10505  & 8.145$\times 10^9$  \newline  2297.97 & 7.084$\times 10^9$ \newline   778.15 & 8.688$\times 10^9$ \newline   246.21\\
\hline
\end{tabular}
\label{enandein2}
\end{table}

\section{Results and discussion}
\label{res}

The interest to the considered problem is triggered by the recent launch of space observatory Spektr-RG on 13 July of 2019. Searching of exotic atoms in Universe was stated as the most substantial part of mission. Within the considered theoretical model we found that the luminescence in the lines of $ \mu p $ and $ (\mu ^4\mathrm{He})^{+} $ atoms can be very noticeable in the spectrum of background source, see Figs. \ref{fig1}-\ref{fig8}. This behavior is similar to the luminescence in lines of primary helium \cite{Dubrovich2018,Dubrovich2018-a}. 

The high efficiency on a wide set of spectral lines could be important for the reliable identification of the origin of spectral lines. In spite of the fact that simple spectral models are used in the present work, the result should not change qualitatively with more accurate models of spectra \cite{shakura}. We also did not consider the question of the specific way in which muonic atoms are formed in the vicinity of astrophysical sources of X-ray emission. All these problems are leaved for future works.

\section*{Acknowledgements}
We thank S. I. Grachev for valuable discussions. T. Z. acknowledges foundation for the advancement of theoretical physics "BASIS".

\end{document}